 \documentclass{emulateapj} 
\shorttitle{Retrograde Orbit for WASP-17b}
\shortauthors{Bayliss et al.}

\begin{document}

\title{Confirmation of a Retrograde Orbit for Exoplanet WASP-17b}

\author{Daniel D.\ R.\ Bayliss\altaffilmark{1}}
\author{Joshua N.\ Winn\altaffilmark{2}}
\author{Rosemary A.\ Mardling\altaffilmark{3}}
\author{Penny D.\ Sackett\altaffilmark{1}}

 \altaffiltext{1}{Research School of Astronomy and
   Astrophysics, The Australian National University, Mt Stromlo
   Observatory, Cotter Rd, Weston Creek, ACT 2611, Australia\\ \email{daniel@mso.anu.edu.au}}
 \altaffiltext{2}{Department of Physics, and Kavli Institute for Astrophysics and Space Research, Massachusetts Institute of Technology, Cambridge, MA 02139}
 \altaffiltext{3}{School of Mathematical Sciences, Monash University, VIC 3800, Australia}

\begin{abstract}
  We present high-precision radial velocity observations of WASP-17
  throughout the transit of its close-in giant planet, using the MIKE
  spectrograph on the 6.5m Magellan Telescope at Las Campanas
  Observatory.  By modeling the Rossiter-McLaughlin effect, we find
  the sky-projected spin-orbit angle to be $\lambda = 167.4\pm
  11.2$~deg.  This independently confirms the previous finding that
  WASP-17b is on a
  retrograde orbit, suggesting it underwent migration via a mechanism
  other than just
  the gravitational interaction between the planet and the disk.
  Interestingly, our result for $\lambda$ differs by $45
  \pm 13$~deg from the previously announced value, and we also find that the
  spectroscopic transit occurs $15\pm 5$~min earlier than 
  expected, based on the published ephemeris.  The discrepancy in the
  ephemeris highlights the need for contemporaneous spectroscopic and
  photometric transit observations whenever possible.
\end{abstract}


\section{Introduction}
\label{sec:introduction}

Knowledge of the angle $\psi$ between a star's spin vector and a
planet's orbital angular momentum vector contains information about the
planetary formation process.  This is especially so for Hot Jupiters,
which are thought to have ``migrated'' inward from large orbital
distances \citep{1996Natur.380..606L}. One possible migration
mechanism involves the gravitational interaction between the planet
and the gas in the protoplanetary disk
\citep[e.g.][]{1996Natur.380..606L,2004ApJ...616..567I}.  Such
interactions would preserve a prograde orbit, and could not account
for Hot Jupiters with retrograde orbits ($\psi>90^\circ$) unless the
protoplanetary disk was initially misaligned with the star
\citep{2010MNRAS.401.1505B}. Other migration mechanisms, involving
planet-planet scattering and Kozai cycles,
\citep[e.g.][]{2008ApJ...678..498N} can produce retrograde orbits.
Therefore by measuring $\psi$ one gains insight into the
migration history of a particular planetary system.

The Rossiter-McLaughlin effect
\citep{1924ApJ....60...15R,1924ApJ....60...22M} occurs when part of
the rotating stellar photosphere is eclipsed by a companion star or
planet.  This removes a velocity component of the observed rotationally-
broadened line profiles, causing a pattern of anomalous radial
velocities to be observed throughout the eclipse.  Although this
effect does not reveal $\psi$, it does give a measure of $\lambda$,
the angle between the {\it sky projection} of the stellar spin vector
and the planetary orbital angular momentum.  The Rossiter-McLaughlin
effect has now been used to measure $\lambda$ for 28 transiting planet
systems (see \citet{2010arXiv1008.2353A} or \citet{2010ApJ...718L.145W}).

The transiting Hot Jupiter
WASP-17b, which was discovered by \citet{2010ApJ...709..159A}, is
reported to have a planetary radius of $R_P=1.74~R_J$ and a host
star rotational velocity of $v\sin i=9$km~s$^{-1}$.  The deep transit
and fast stellar rotation make it a propitious target for
Rossiter-McLaughlin effect observations.

\citet{2010ApJ...709..159A} reported that the planet appeared to be in
a retrograde orbit, based on three radial-velocity measurements that
were obtained during transits.  Subsequently, the same group announced
further radial velocity measurements of WASP-17, with denser time
sampling \citep{2010arXiv1008.2353A}. Those data gave much stronger support to the
claim of a retrograde orbit, and indeed gave a value for $\lambda$
nearly identical to the value derived by
\citet{2010ApJ...709..159A}. In this Letter we present new
observations and an independent confirmation of the likely retrograde orbit of the WASP-17 system, though with a significantly different derived value for $\lambda$.

\section{Observations}
\label{sec:observations}

On the night of 2010 May 11, we obtained high-resolution spectra of WASP-17 using
the Magellan Inamori Kyocera Echelle (MIKE) spectrograph on the 6.5m
Magellan II (Clay) telescope at Las Campanas Observatory.  MIKE was
used in its standard configuration with a 0.35\arcsec~slit and fast
readout mode, which delivered 32 red echelle orders with wavelengths
from 5000-9500~\AA~and with a resolving power of $R\approx48,000$.
Spectra were taken continuously for 9 hours, with an interruption of
approximately 30 minutes when the star crossed the meridian near the zenith.  A total
of 33 spectra of WASP-17 were obtained, with 600~s exposures.
ThAr calibration frames bracketed each spectrum. Observing conditions were
good, although some high, thin clouds were present in the latter half
of the night.  The moon was down throughout the WASP-17 observations,
and the seeing was 0.5-0.7\arcsec.

The spectra were reduced using the Carnegie MIKE pipeline developed by
Dan Kelson.  Radial velocities were determined via cross correlation
with respect to a template spectrum.  We tested different choices for
the template spectrum, including a radial velocity standard observed
on the same night and various individual WASP-17 spectra. The
lowest inter-order scatter was obtained when using the highest
signal-to-noise WASP-17 spectrum as the template. This spectrum was
also obtained at low airmass (1.005).  For the cross-correlation
analysis we selected 13 echelle orders that were free of obvious
telluric lines and had a signal-to-noise ratio exceeding
100~pixel$^{-1}$. Cross-correlation was performed using the
IRAF\footnote{IRAF is distributed by the National Optical Astronomy
  Observatories, which are operated by the Association of Universities
  for Research in Astronomy, Inc., under cooperative agreement with
  the National Science Foundation.} task $fxcor$.  We used the O$_{2}$
absorption features from 6870-6900~\AA~to provide a constant
reference for our wavelength solutions for each spectrum.  These
absorption bands have been shown to be stable at the 5~m~s$^{-1}$
level over short timescales \citep{2010arXiv1003.0541F}.

MIKE lacks an atmospheric dispersion corrector, and this resulted in an
airmass-dependent systematic trend in the radial velocity
measurements, with an amplitude that increased for bluer
echelle orders. The continuous nine-hour monitoring allowed us to
track this systematic trend and account for it during the
model-fitting procedure, as described in Section~\ref{sec:analysis}.
The order-averaged, airmass-corrected radial velocities are given in Table~1, with
uncertainties taken to be the standard deviation of the mean of the
results of all 13 orders.

\section{Analysis}
\label{sec:analysis}

Our model for the radial velocity data has the form:
\begin{equation}
  V_{{\rm calc}, n}(t) = V_{\rm orb}(t) + V_{\rm RM}(t) + c_0 + c_1n +
  (c_2 + c_3n)X,
\end{equation}
where $V_{{\rm calc},n}(t)$ is the calculated radial velocity at time
$t$ in echelle order $n$ (ranging from 1 to 13), $V_{\rm orb}$ is the
radial velocity due to the star's orbital motion (assumed to be
circular), $V_{\rm RM}$ is the transit-specific ``anomalous velocity''
due to the Rossiter-McLaughlin effect, and $\{ c_0, c_1, c_2, c_3\} $
are constants specifying the offset between the relative barycentric
velocity of WASP-17 and the arbitrary template spectrum that was used
to calculate radial velocities. To account for the
wavelength-dependent airmass trend mentioned in the previous section,
the offset was allowed to be a linear function of both order number
$n$ (effectively a wavelength index) and airmass $X$.

Many approaches have been taken to model the Rossiter-McLaughlin effect, such as the
pixellated photospheric model of \citet{2000A&A...359L..13Q}, the
first-moment approach of \citet{2005ApJ...622.1118O} and
\citet{2006ApJ...650..408G}, the forward-modeling approach of
\citet{2005ApJ...631.1215W} or the spectral line fitting of
\citet{2009Natur.461..373A} and \citet{2010MNRAS.403..151C}. In our
case the most applicable and convenient model is the analytic formula
of \citet{2010ApJ...709..458H},
\begin{eqnarray}
  V_{\rm RM}(t) =-\delta(t)~v_p(t) \left[
\frac{2v_0^2 + 2(v\sin i)^2}{2v_0^2 + (v\sin i)^2}
\right]^{3/2}\nonumber \\
\left[
1 - \frac{v_p(t)^2}{2v_0^2 + (v\sin i)^2}
\right],
\end{eqnarray}

where $\delta(t)$ is the loss of light during the transit, $v_p(t)$ is
the mean radial velocity of the small portion of the photosphere that
is hidden by the planet, $v\sin i$ is the stellar projected rotation
rate, and $v_0$ is the velocity width of the spectral lines that would
be seen from a small portion of the photosphere (i.e.\ due to
mechanisms other than rotation). This formula relates the phase of the
transit to the radial velocity anomaly derived from cross-correlation.

To calculate $v_p(t)$ we assumed that the stellar photosphere rotates
uniformly with an angle $\lambda$ between the sky projections of the
spin vector and the orbital angular momentum vector (see, e.g.,
\citet{2005ApJ...622.1118O} or \citet{2007ApJ...655..550G}). We took
$v\sin i$ to be a free parameter, and $v_0$ was constrained by the
condition that the quadrature sum of $v_0$ and $v\sin i$ should equal
the observed total linewidth of 11~km~s$^{-1}$
\citep{2010ApJ...709..159A}.

\begin{figure}[tpb]
\epsscale{1.0}
\plotone{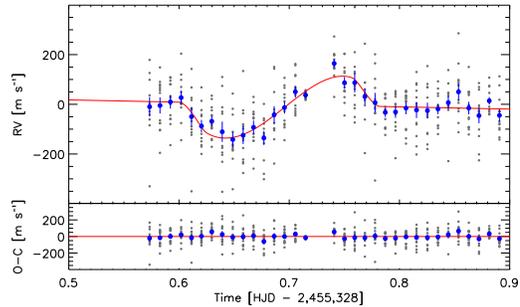}
\caption{ {\bf The spectroscopic transit of WASP-17b.}
  {\it Top.}---Radial velocity variation observed on the night of 2010 May 11-12,
  along with the best-fitting model. The small gray points are the
  order-by-order radial velocities that were fitted. The larger blue points are
  averages of the results from all orders.
  {\it Bottom.}---Residuals between the data and the best-fitting model.
  \label{fig:rv}}
\end{figure}

To calculate $\delta(t)$, we assumed a linear limb-darkening law with
a fixed coefficient of 0.7, and used the analytic formulas of
\citet{2002ApJ...580L.171M}, as implemented by
\citet{2008MNRAS.390..281P}.  The parameters of the photometric model
were the planet-to-star radius ratio $R_p/R_\star$, orbital
inclination $i$, and normalized stellar radius $R_\star/a$ (where $a$
is the orbital distance).  Some of the transit characteristics are
more tightly constrained by previous observations of photometric transits
than by the radial velocity data presented here.  Hence, we used
Gaussian priors for the transit depth $(R_p/R_\star)^2$, total
duration $t_{\rm IV}-t_{\rm I}$, partial duration $t_{\rm IV}-t_{\rm
  III}$, orbital period $P$, and time of transit $T_c$, based on the
results presented by \citet{2010ApJ...709..159A} (see their Table 4,
case 3). We also used those results to set a Gaussian prior for $K$,
the orbital velocity semi-amplitude, since our coverage of the
spectroscopic orbit is sparse.

All together there are 11 adjustable parameters in our model fit, of which 7 are 
mainly determined by our new radial velocity data: $\lambda$, $v\sin
i$, $c_0$, $c_1$, $c_2$, $c_3$, and the time of conjunction $T_c$. The
other 4 parameters, $R_p/R_\star$, $i$, $R_\star/a$, and $K$, are 
controlled mainly by the priors. We determined the best-fitting
parameter values and their 68.3\% confidence limits with a Monte Carlo
Markov Chain (MCMC) algorithm that we have described elsewhere (see,
e.g., \citet{2007AJ....133...11W}). The likelihood was taken to be
$\exp(-\chi^2/2)$ with
\begin{equation}
\chi^2 =
\sum_{i=1}^{33}
\sum_{n=1}^{13}
  \left[ \frac{v_{\rm obs}(t_i, n) - v_{\rm calc}(t_i, n)} {\sigma} \right]^2,
\end{equation}
where $v_{\rm obs}(t_i, n)$ is the radial velocity measured at time
$t_i$ in echelle order $n$, and $v_{\rm calc}(t_i, n)$ are the
calculated radial velocities. We found that the choice
$\sigma=92$~m~s$^{-1}$ gave $\chi^2=N_{\rm dof}$ for the best-fitting
model, and is also approximately equal to the scatter observed in the
residuals to the best-fitting model, so we used this value in the MCMC
analysis.

\begin{figure}[tpb]
\epsscale{0.8}
\plotone{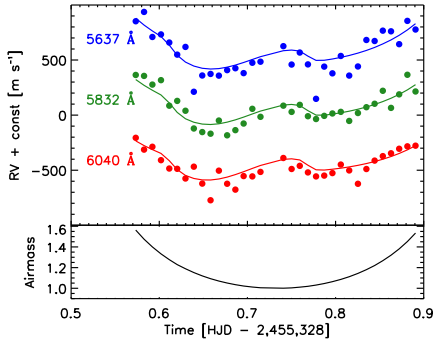}
\caption{ {\bf Illustration of airmass correction.}  The
  radial velocities for echelle orders 61, 59, and 57 (from top to bottom), along with
  the best-fitting model that includes both an airmass correction and the Rossiter-McLaughlin effect.
  Central wavelengths of the orders are indicated on the plot.
  The lower panel shows the time dependence of the airmass.
  \label{fig:airmass}}
\end{figure}

\begin{figure}[tpb]
\epsscale{1.0}
\plotone{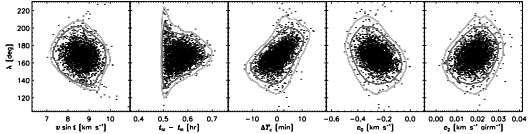}
\caption{ {\bf Results for the model parameters,} based on our
   MCMC analysis of the radial velocity data, for parameters which
   show a degree of correlation with $\lambda$. The contours represent 68\%, 95\%, and
  99.73\% confidence limits. The quantity $\Delta T_c$ is defined as
  $T_c$ minus the optimized value of HJD 2,455,328.6918.
\label{fig:mcmc}}
\end{figure}

Table~\ref{tbl:params} gives the resulting best-fit parameter
values. Figure~\ref{fig:rv} shows the airmass-corrected radial
velocities as a function of time, along with the best-fitting model
and the residuals.  Figure~\ref{fig:airmass} illustrates the airmass
correction for three separate echelle orders.  Figure~\ref{fig:mcmc}
displays the marginalized, joint {\it a posteriori} probability
distributions involving $\lambda$ and the other model parameters.

The radial velocity data clearly exhibit an ``upside-down''
Rossiter-McLaughlin effect: an anomalous blueshift for the first half
of the transit, followed by an anomalous redshift for the second
half. Our result for the projected stellar rotation velocity, $v\sin
i_\star=8.61\pm 0.45$~km~s$^{-1}$, is in agreement with the results of
\citet{2010ApJ...709..159A} and \citet{2010arXiv1008.2353A}.  However, our finding
of $\lambda=167.4\pm 11.2$~deg does not agree with either of the
previously reported results.\footnote{For ease of comparison, we added
  360~deg to the result quoted by \citet{2010ApJ...709..159A}, and we
  converted the result of \citet{2010arXiv1008.2353A} onto our coordinate system
  using $\lambda = -\beta$.} \citet{2010ApJ...709..159A} found
$\lambda=210.7^{+11.5}_{-8.9}$~deg, although this result was based on
only 3 data points. More significant is the result by \citet{2010arXiv1008.2353A}
of $\lambda=212.7^{+5.5}_{-5.9}$~deg, based on two data sets densely
sampling the transit.  The difference between our result for $\lambda$
and that of \citet{2010arXiv1008.2353A} is $45 \pm 13$~deg, or 3.5$\sigma$ away
from zero.  

We do not know the reason for the discrepancy. We have sparser
coverage of the spectroscopic orbit than did
\citet{2010arXiv1008.2353A}, but this did not bias our results: we
used the results of \citet{2010ApJ...709..159A} to set a prior for
$K$, and also found that the results for $K$ and $\lambda$ are
uncorrelated. We also checked to see if the gap in our time coverage
when WASP-17 was at zenith has biased our results: we fitted fake data
with $\lambda=210\fdg 7$ and the same time coverage and velocity
errors as our MIKE data, and found $\lambda = 206^\circ \pm 9^\circ$,
i.e., the result was not biased. We also fitted the
\citet{2010arXiv1008.2353A} data ourselves and found $\lambda =
208^\circ \pm 12^\circ$.  Although this reduces the discrepancy in
results to 2.5$\sigma$, it indicates that the differences in $\lambda$
values cannot be attributed solely to differences in fitting
procedures.  An important step in our data analysis was fitting and
removing the airmass trend in our radial velocity data.  We fitted for the airmass terms simultaneously with the Rossiter-McLaughlin
model, and therefore our results do take into account any correlations
between the airmass parameters and the other parameters including
$\lambda$. However, these errors are internal to the choice of our
model, which is linear in airmass and order number. This model seems
reasonable and provides a satisfactory fit to the data; however it is
impossible to exclude the possibility that the true dependence is more
subtle and that this has altered the shape of the Rossiter-McLaughlin effect.

According to the ephemeris of \citet{2010ApJ...709..159A}, the
predicted time of conjunction for the event we observed is
HJD~2,455,328.6814$~\pm 0.0017$, which is earlier by $15\pm 5$~min
than the time we measured. Either the transits are nonperiodic, or the
uncertainties in at least some of the transit times were
underestimated. A straight-line fit to the epochs and transit times in
Table~5 of \citet{2010ApJ...709..159A} gives $\chi^2=36$ with 11
degrees of freedom, i.e., statistically different than random. For this reason,
it is likely that the true uncertainty in the period is larger than
the previously reported uncertainty, by a factor of about
$\sqrt{36/11}$ or 1.8.

Consequently, we conclude that our timing offset from the ephemeris of
\citet{2010ApJ...709..159A} is at most a 1.5$\sigma$ discrepancy.  The
uncertainty in the time of conjunction would be best addressed in the future
by obtaining high-quality photometric data simultaneously with
spectroscopic observations.  However we note that this ephemeris discrepancy
cannot account for the difference between our result for $\lambda$ and that 
of \citet{2010arXiv1008.2353A}.

Our result for $\lambda$ is consistent with a retrograde orbit, but it
must be remembered that $\lambda$ is a sky-projected quantity. The
true angle $\psi$ between the vectors is given by
\begin{equation}
\cos\psi = \cos i_\star \cos i + \sin i_\star \sin i \cos\lambda,
\end{equation}
where $i$ and $i_\star$ are the line-of-sight inclinations of the
orbital and stellar angular momentum vectors, respectively.  Using
$i=86.95_{-0.63}^{+0.87}$~deg from \citet{2010ApJ...709..159A}, and
supposing $i_\star$ to be drawn from an ``isotropic'' distribution
(uniform in $\cos i_\star$), we find $\psi > 92\fdg 6$ with 99.73\%
confidence. In this sense, the WASP-17b orbit is very likely to be
retrograde ($\psi > 90^\circ$), although nearly pole-on configurations
($\psi \approx 90^\circ$) are possible.

The assumption of an isotropic distribution in $i_\star$ neglects our
prior knowledge that main-sequence stars have somewhat predictable
rotation rates, and therefore that the measured value of $v\sin
i_\star$ also bears information about $i_\star$.  Recently,
\citet{2010arXiv1006.2851S} used this insight to seek evidence for
small values of $\sin i_\star$ (and therefore large spin-orbit
misalignments along the line of sight) among all the transit hosts. He
did not identify WASP-17 as one of the systems with a likely small
value of $\sin i_\star$. In light of this analysis, our observation of
a large $\lambda$ favors a more nearly retrograde orbits over polar
orbits.

\section{Discussion}
\label{sec:discussion}

It is now ten years since the first reported detection of the
Rossiter-McLaughlin effect due to a transiting planet
\citep{2000A&A...359L..13Q}.  As with many other properties of
extrasolar planets, the Rossiter-McLaughlin effect observations have
thrown up surprises, with systems that appear very different from
those in our own Solar System.  The Hot Jupiter HAT-P-7b was found to
be on a polar or retrograde orbit, by independent measurements of
\citet{2009PASJ...61L..35N} and \citet{2009ApJ...703L..99W}.
Similarly WASP-17b has now been confirmed to be in a retrograde orbit
by two independent groups \citep[][this work]{2010arXiv1008.2353A}.  It now seems
that at least a fraction of Hot Jupiters are migrating by mechanisms
other than disk migration, or that proto-planetary disks are
frequently misaligned with their stars.

The work of \citet{2009ApJ...696.1230F} suggested that we may be
seeing two different populations of Hot Jupiters that have migrated
via different mechanisms.  The results of this work adds weight to
that suggestion, however a larger statistical sample will be needed
before this can be robustly confirmed.  Only a larger
sample of planets with known $\lambda$ values will allow correlations
with parameters such as planetary mass, planetary radius, metallicity,
and stellar mass to be robustly tested.  
\citet{2010ApJ...718L.145W} have proposed that misaligned Hot Jupiters
occur preferentially around hot stars ($T_{\rm eff} > 6250$~K), of
which WASP-17 is a supporting example. Fortunately the discovery rate
of transiting planets orbiting bright stars is set to rise due to
large-scale ground-based surveys such as SuperWASP
\citet{2007MNRAS.375..951C} and HAT-South \citep{2009IAUS..253..354B},
and proposed space-based projects such as TESS
\citep{2009PASP..121..952D} and PLATO \citep{2008JPhCS.118a2040C}.

\acknowledgments Australian access to the Magellan Telescopes was
supported through the National Collaborative Research Infrastructure
Strategy of the Australian Federal Government.  DDRB and RAM acknowledge
financial support from the Access to Major Research Facilities
Programme, which is a component of the International Science Linkages
Programme established under the Australian Government’s innovation
statement, Backing Australia’s Ability.  JNW gratefully acknowledges support from the
NASA Origins program through award NNX09AD36G, and from a generous
gift of the MIT Class of 1942.  We thank Amaury Triaud for providing
the radial velocity data for WASP-17 presented by \citet{2010arXiv1008.2353A}.  DDRB and RAM would also like
to thank St\'{e}phane Udry for his helpful discussions on subjects
relating to this work. 

{\it Facilities:} \facility{Magellan: Clay (MIKE)}

\bibliographystyle{apj}

\clearpage

\begin{deluxetable}{rcc}

\tablecaption{WASP-17 Radial Velocities (order averaged)\label{tbl:rv}}
\tablewidth{0pt}
\tablehead{
\colhead{Time} & \colhead{Radial}  & \colhead{Radial velocity} \\
\colhead{HJD} & \colhead{velocity (m~s$^{-1}$)} & \colhead{uncertainty (m~s$^{-1}$)}
}
\startdata
$2455328.57331$ & $  -9.88$ & $  38.75$ \\
$2455328.58273$ & $  -4.73$ & $  32.47$ \\
$2455328.59219$ & $   9.70$ & $  27.08$ \\
$2455328.60194$ & $  26.88$ & $  27.51$ \\
$2455328.61135$ & $ -49.67$ & $  37.43$ \\
$2455328.62035$ & $ -87.48$ & $  24.96$ \\
$2455328.62971$ & $ -68.18$ & $  20.08$ \\
$2455328.63921$ & $-110.63$ & $  40.87$ \\
$2455328.64876$ & $-141.71$ & $  30.59$ \\
$2455328.65817$ & $-124.57$ & $  35.23$ \\
$2455328.66761$ & $ -92.68$ & $  27.30$ \\
$2455328.67700$ & $-135.62$ & $  27.72$ \\
$2455328.68633$ & $ -42.40$ & $  34.15$ \\
$2455328.69570$ & $ -13.34$ & $  24.05$ \\
$2455328.70557$ & $  50.30$ & $  23.05$ \\
$2455328.71481$ & $  37.16$ & $  15.00$ \\
$2455328.74077$ & $ 164.59$ & $  17.62$ \\
$2455328.75012$ & $  86.38$ & $  23.37$ \\
$2455328.75939$ & $  86.77$ & $  38.93$ \\
$2455328.76863$ & $  31.27$ & $  30.88$ \\
$2455328.77785$ & $   6.61$ & $  39.38$ \\
$2455328.78711$ & $ -32.55$ & $  18.04$ \\
$2455328.79641$ & $ -30.80$ & $  24.08$ \\
$2455328.80593$ & $ -15.31$ & $  24.74$ \\
$2455328.81546$ & $ -23.01$ & $  32.34$ \\
$2455328.82536$ & $ -25.39$ & $  32.21$ \\
$2455328.83495$ & $ -18.66$ & $  20.94$ \\
$2455328.84430$ & $   6.76$ & $  26.41$ \\
$2455328.85358$ & $  50.02$ & $  32.97$ \\
$2455328.86289$ & $ -15.15$ & $  25.22$ \\
$2455328.87214$ & $ -45.57$ & $  30.49$ \\
$2455328.88145$ & $  13.95$ & $  15.26$ \\
$2455328.89069$ & $ -44.50$ & $  24.20$
\enddata

\end{deluxetable}

\clearpage

\begin{deluxetable}{lccccc}

\tabletypesize{\footnotesize}
\tablecaption{Model Parameters for WASP-17b\label{tbl:params}}
\tablewidth{0pt}

\tablehead{
\colhead{Parameter} &
\colhead{Value}
}

\startdata
\hline
{\it Parameters determined mainly by the}\\
{\it new radial velocity data} \\ 
\hline
Projected spin-orbit angle, $\lambda$~[deg]                       & $167.6\pm 11.4$ \\ 
Projected stellar rotation rate, $v\sin i_\star$~[km~s$^{-1}$] & $8.61\pm 0.45$ \\
Radial velocity offset, $c_0$ [m~s$^{-1}$]                            & $-0.267 \pm 0.074$ \\
Order dependence of radial velocity offset, $c_1$ [m~s$^{-1}$~order$^{-1}$]  & $-0.79 \pm 0.11$ \\
Airmass term, $c_2$ [m~s$^{-1}$~airmass$^{-1}$]  & $0.233\pm 0.061$ \\
Order dependence of airmass term, $c_3$ [m~s$^{-1}$~airmass$^{-1}$~order$^{-1}$]  & $0.738\pm 0.093$ \\
\hline 
{\it Parameters determined mainly by priors} \\
\hline  
Velocity semi-amplitude, $K$~[m~s$^{-1}$]            & $55.9\pm 5.0$ \\ 
Time of conjunction, $T_c$~[HJD-2,450,000]              & $5,328.6919 \pm 0.0027$  \\ 
Transit depth, $(R_p/R_\star)^2$                    & $0.01676 \pm 0.00027$ \\
Total transit duration, $t_{\rm IV}-t_{\rm I}$~[hr]      & $4.368 \pm 0.037$ \\
Partial transit duration, $t_{\rm IV}-t_{\rm III}$~[hr]      & $0.566 \pm 0.039$\\
\hline
\enddata
\end{deluxetable}


\end{document}